\documentclass[amsmath,amssymb,twocolumn]{revtex4}
\usepackage[parfill]{parskip}    
\usepackage{graphicx}
\usepackage{amssymb}
\usepackage{epstopdf}
\usepackage{color}
\usepackage{graphicx}
\usepackage{pdfpages}
\usepackage[utf8]{inputenc}

\begin{document}

\title{Gravitational Decoupling in Cosmology}
\author{Francisco X. Linares Cede\~no\footnote{francisco.linares@umich.mx}}
\address{Instituto de F\'isica y Matem\'aticas, Universidad Michoacana de San Nicol\'as de Hidalgo, Edificio C-3, Ciudad Universitaria, CP. 58040 Morelia, Michoac\'an, M\'exico.}
\author{Ernesto Contreras {\footnote{econtreras@usfq.edu.ec}}}
\address{ Departamento de F\'isica, Colegio de Ciencias e Ingenier\'ia, Universidad San Francisco de Quito, Quito, Ecuador\\}

\begin{abstract}

Whereas the nature of dark components in the Universe remains unknown, alternative models of gravity have been developed to offer a geometric explanation to the origin of such components. In this work we use the Minimal Geometric Deformation approach to study extensions of the theory of General Relativity in a cosmological context. This is possible since such approach allows the decoupling of gravitational sources, and the Einstein field equations can be analytically solved with the presence of a new gravitational sector once a known GR solution is considered. In particular, we implement such approach in Friedmann-Robertson-Walker and Kantowski-Sachs universes. We demonstrate that the gravitational decoupling leads to modifications of well known cosmological solutions. For instance, we show that an effective spatial curvature in the Friedmann-Robertson-Walker metric, as well as several kind of matter components in the Kantowski-Sachs case, are obtained. Thus, we found that it is possible to obtain spatial curvature and new matter terms from geometry, which in cosmology they could be useful in addressing problems such as the spatial flatness of the Universe, dark matter and dark energy.
\end{abstract}

\maketitle

\section{Introduction}\label{intro}

Current cosmological and astrophysical observations indicate that the most accepted cosmological model is the so-called $\Lambda$CDM~\cite{Eisenstein:2005su,WiggleZPhysRevD86,Blake2012MNRAS,Kazin:2014qga,Beutler:2015tla,Ade:2015xua,Aghanim:2018eyx}, which offers an accurate phenomenological description of the evolution of the Universe. According to this model, only $\sim 4\%$ of total matter content is constituted by ordinary matter made of the known fundamental particles. Another $\sim 26\%$ is attributed to Cold Dark Matter (CDM), a non-relativistic particle whose interaction is mostly gravitational, whereas the remaining $\sim 70\%$ belongs to Dark Energy (cosmological constant $\Lambda$), responsible of the current accelerated expansion of the Universe. This model is based on the mathematical framework of \textit{General Relativity} (GR), which has shown to be a successful theory of gravity~\cite{Will:2014kxa,Will:2018lln}. However, GR by itself does not predict the existence of these new and enigmatic components of our Universe. In the case of dark matter, it is usually introduced by hand in the \textit{Einstein Field Equations} (EFE) as a new matter component (new particles described in terms of some energy-momentum tensor), whereas for dark energy, physical and mathematical principles leading to modifications of the law of gravity (new geometric terms correcting the Einstein tensor) have been proposed. Thus, despite the success of the $\Lambda$CDM model, the true nature of dark matter and dark energy remains unknown.

This fact has motivated many alternative models. Particularly, there are theoretical proposals in which the law of gravity changes at large scales, such as \textit{MOND} (\textit{MO}dified \textit{N}ewtonian \textit{D}ynamics), where the acceleration at galactic scale obeys a different law of gravitation~\cite{milgrom1983modification,milgrom2002mond,milgrom2014mond,sanders2002modified,corbelli2007testing}, $f(R)$ theories, where higher order terms of the Ricci scalar modify the equations of motion~\cite{sotiriou2010f,cembranos2009dark,he2015effective,bohmer2008dark,capozziello2006dark}, or \textit{Braneworld} models, where the EFE are generalized with new tensors arising from an extra spatial dimension~\cite{maartens2010brane,germani2001stars,cembranos2003brane,garcia2012braneworld,okada2004relic}. There are also some proposals based on what is called \textit{Modified Gravity}, in which geometric extensions of the theory of GR are proposed to explain the late time acceleration~\cite{Nojiri:2006ri,Lobo:2008sg,Nojiri:2010wj,Tsujikawa:2010sc,Capozziello:2010zz,Bamba:2012cp,Joyce:2014kja,Koyama:2015vza,Li:2011sd,Capozziello:2011et,Clifton:2011jh,Dimitrijevic:2012kb,Brax:2015cla,Joyce:2016vqv,Jaime:2018ftn,deAlmeida:2018kwq}, and which have been applied to the realm of dark matter as well~\cite{Choudhury:2015zlc,Aoki:2016zgp,Katsuragawa:2016yir,Katsuragawa:2017wge,Shi:2017pyd}.

Such extensions of GR lead to a more complicated set of equations to be solved. In fact, it is well known that due to the non-linearity of the EFE, it is in general a difficult task to obtain new and physical solutions in GR, even for static and spherically symmetric spacetimes. Moreover, if a set of solutions is known, it is not true that a linear combination of them leads to new solutions of the EFE, this is, the superposition principle is not valid in GR. In spite of such complexity, the \textit{Minimal Geometric Deformation approach} (MGD)~\cite{Ovalle:2007bn,Ovalle:2008se,Ovalle:2010zc} have been consolidated as a powerful and efficient way to study the decoupling of gravitational sources, in presence of modifications of gravity with the form~\cite{Ovalle:2017khx,ovalle2020minimal}
\begin{equation}
    S = \int d^4x \sqrt{-g}\left(\frac{R}{2\kappa^2} + \mathcal{L}_m\right) + \alpha \left({\rm{correction}}\right)\, , 
\end{equation}
where $\kappa^2=8\pi G$, and $\alpha$ is a free parameter associated with the correction due to the new gravitational sector. Specifically, the MGD approach allows to find solutions of the EFE in presence of a new gravitational source, by using a known solution. This have been verified in several problems in the context of relativistic astrophysics~\cite{Casadio:2012pu,Ovalle:2013xla,Ovalle:2013vna,Ovalle:2014uwa,Casadio:2015jva,Casadio:2015gea,Ovalle:2015nfa,Cavalcanti:2016mbe,Casadio:2016aum,daRocha:2017cxu,daRocha:2017lqj,Casadio:2017sze,Morales:2018urp,Morales:2018nmq,Panotopoulos:2018law,Estrada:2018vrl,Maurya:2019wsk,Heras:2019ibr,Contreras:2019mhf,Gabbanelli:2019txr,Maurya:2019hds,Sharif:2019rln,Hensh:2019rtb}. The first application in the context of GR was considered in~\cite{Ovalle:2017fgl} to extend a Tolman IV solution to anisotropic domains. After this work the method has been implemented not only to extend interior solutions~\cite{Estrada:2018zbh,Heras:2018cpz,Gabbanelli:2018bhs,Sharif:2018toc,Sharif:2018tiz,Fernandes-Silva:2018abr}, but to decouple the matter sector of black holes in $(2+1)$~\cite{Contreras:2018vph,Contreras:2019fbk} and $(3+1)$~\cite{Ovalle:2018umz} dimensional spacetimes and gravastars~\cite{Ovalle:2019lbs}. Furthermore, it has been shown that the MGD provides a framework to study any GR modification associated with a conformal gravitational sector, as is the case for braneworld and $f(R)$ models~\cite{Ovalle:2017khx,ovalle2020minimal}.

We want to emphasize that the decoupling of gravitational sources is a highly non-trivial theoretical problem. In this sense, the MGD is not just a technique to solve the EFE. Moreover, the power of this formalism lies in the fact that it allows to solve in a direct and systematic way, the problem of decoupling gravitational sources in GR. In particular, it is possible to induce a new matter term by means of the introduction of geometric deformations on the metric tensor, and finding the solution for this new term in a consistent way.

While the MGD approach have been focused so far on astrophysical systems, the idea of having new matter terms arising from geometry motivate us to explore the consequences of this approach on cosmological scenarios. There are matter components in the Universe still unknown, and our main motivation is to analyse the cosmic fluids arising as consequence of geometric considerations in the framework of the MGD. Thus, it is our main goal here to use the MGD approach in order to extend well-known cosmological solutions, to new ones including a new gravitational source induced by a geometric deformation on the metric. In this sense, we will not consider a particular theory of modified gravity, but we will take the simplest generic way to extend GR through the MGD. The protocol presented in this work could be important in cosmology since, as we will see, it gives us clues about a possible geometric origin of the dark components present in our Universe.

This work is organized as follows. In Section~\ref{review} we review the basics of the MGD approach, and explain how it allows to decouple gravitational sources. Section \ref{FRW} is devoted to the reformulation of the MGD in a Friedmann-Robertson-Walker geometry. We adapt the MGD to extend solutions of 
Kantowski-Sachs spacetimes in Section \ref{KS}. The summary and perspectives of the work are given in the last Section.

\section{MGD and gravitational decoupling: a brief review}\label{review}

Let us explain the MGD approach, and how it can be used to decouple gravitational sources. Suppose that certain well known solution of the EFE has a line element parameterized as
\begin{eqnarray}\label{lei}
ds^{2}=-e^{\nu(r)}dt^{2}+\frac{dr^{2}}{\mu(r)}+r^{2}d\Omega^{2},
\end{eqnarray}
in presence of a perfect fluid $T^{\mu}_{\nu}=diag(-\rho,p,p,p)$, and where the gravitational potentials $\mu$ and $\nu$ are functions only of the radial coordinate $r$. Now, in order to extend the isotropic solution to anisotropic domains by means of the MGD, we consider the following transformations~\cite{Ovalle:2017fgl}:
\begin{subequations}\label{gdt}
\begin{eqnarray}
\nu \rightarrow \xi = \nu + \alpha g\, , \\
\mu \rightarrow e^{-\lambda} = \mu + \alpha f\, , 
\end{eqnarray}
\end{subequations}
where again, $\alpha$ is a free parameter with constant value measuring the strength of the geometric deformation induced in the gravitational potentials $\nu$ and $\mu$ by the decoupling functions $g$ and $f$. From all the possible transformations given by Eq.~\eqref{gdt}, there is the so-called \textit{minimal geometric deformation}, for which $g=0$ and $f\neq 0$. Thus, the transformation will lie only in the radial component.

Now, we implement the following protocol: first, we introduce a deformation in the $g^{rr}$ component of the metric in the following way
\begin{eqnarray}\label{decomposition}
\mu(r)\to e^{-\lambda(r)}=\mu(r)+\alpha f(r),
\end{eqnarray}

Then, we have to consider a more general energy-momentum tensor $T_{\mu\nu}^{tot}$ given by
\begin{eqnarray}\label{total}
T_{\mu\nu}^{tot}=T_{\mu\nu}
+\alpha\theta_{\mu\nu},
\end{eqnarray}
where $T^{\mu}_{\nu}$ is the perfect fluid mentioned before, and $\theta^{\mu}_{\nu}=diag(-\rho^{\theta},p_{r}^{\theta},p_{\perp}^{\theta},p_{\perp}^{\theta})$ is the anisotropic sector induced by the decoupling function $f$. Notice that if we define $\tilde{\rho}\, ,\tilde{p}_r\, ,\tilde{p}_{\perp}$ as the components of the total energy-momentum tensor, then Eq.~{\eqref{total}} can be written as
\begin{subequations}\label{totalemt}
\begin{eqnarray}
\tilde{\rho}&=&\rho+\alpha \rho^{\theta}\, ,\\
\tilde{p}_{r}&=&p+\alpha p_{r}^{\theta}\, ,\\
\tilde{p}_{\perp}&=&p+\alpha p_{\perp}^{\theta}\, .
\end{eqnarray}
\end{subequations}

Finally, we impose that the line element including the deformation~\eqref{decomposition} given by
\begin{eqnarray}\label{lel}
ds^{2}=-e^{\nu}dt^{2}+e^{\lambda}dr^{2}+r^{2}d\Omega^{2},
\end{eqnarray}
 is a solution of the EFE,
\begin{eqnarray}\label{einsorig}
R_{\mu\nu}-\frac{1}{2}R g_{\mu\nu}=\kappa^{2}T_{\mu\nu}^{tot}.
\end{eqnarray}

What follows is to compare terms. After some algebraic computations, we obtain two sets of differential equations, one for the perfect fluid
\begin{subequations}\label{isogrin}
\begin{eqnarray}
\rho &=&- \frac{r \mu '+\mu -1}{\kappa ^2 r^2}\, ,\label{iso1}\\
p &=& \frac{r \mu \nu '+\mu -1}{\kappa ^2 r^2}\, ,\label{iso2}\\
p &=& \frac{r \mu ' \nu '+2 \mu '+2 r \mu  \nu ''+r \mu  \nu '^2+2 \mu \nu '}{4 \kappa ^2 r}
,\label{iso3}
\end{eqnarray}
\end{subequations}
and a second one for the anisotropic sector,
\begin{subequations}\label{anisomgdin}
\begin{eqnarray}
\rho^{\theta} &=& -\frac{r f'+f}{\kappa ^2 r^2}\, ,\label{aniso1}\\
p_{r}^{\theta}&=& \frac{r f \nu '+f}{\kappa ^2 r^2}\, ,\label{aniso2}\\
p_{\perp}^{\theta}&=&\frac{r f' \nu '+2 f'+2 r f \nu ''+r f \nu '^2+2 f \nu '}{4 \kappa ^2 r},\label{aniso3}
\end{eqnarray}
\end{subequations}
where primes indicate derivative with respect to the radial coordinate $r$.

It is worth recalling that, since the Einstein tensor is divergence free, the total energy-momentum tensor must satisfy $\nabla_{\nu}T^{(tot)\mu \nu}=0\, .$ An explicit computation reveals that this leads to
\begin{equation}\label{constotoexpli}
p^{\prime}+\frac{\nu^{\prime}}{2}(\rho+p)+
\alpha \left[ p_{r}^{\theta \prime}+\frac{\nu^{\prime}}{2}(\rho^{\theta}+p_{r}^{\theta})+ \frac{2}{r}(p_{\perp}^{\theta}-p_{r}^{\theta})\right]=0\, ,
\end{equation}
from which we can read the $\alpha$--independent term as the conservation of the energy-momentum tensor, this is
\begin{equation}\label{conspf}
\nabla_{\mu}T^{\mu}_{\nu}=p^{\prime}+\frac{\nu^{\prime}}{2}(\rho+p)=0\, ,
\end{equation}
and thus, the conservation of $\theta_{\mu\nu}$ can be written as
\begin{eqnarray}\label{constheta}
\nabla_{\mu}\theta^{\mu}_{\nu}=p_{r}^{\theta \prime}+\frac{\nu^{\prime}}{2}(\rho^{\theta}+p_{r}^{\theta})+ \frac{2}{r}(p_{\perp}^{\theta}-p_{r}^{\theta})=0\, .
\end{eqnarray}

After comparing Eqs. (\ref{constotoexpli}), (\ref{conspf}) and (\ref{constheta}) we observe that the perfect fluid $T^{\mu}_{\nu}$ and the anisotropic sector $\theta^{\mu}_{\nu}$ interact only gravitationally \cite{Ovalle:2017wqi,Ovalle:2017fgl,Ovalle:2018umz}, this is
\begin{equation}
\nabla_{\mu}T^{\mu}_{\nu}=\nabla_{\mu}\theta^{\mu}_{\nu}=0\, .
\label{conseq}
\end{equation}

For a given solution of Eq.~\eqref{isogrin} for the gravitational potentials $\{\nu,\mu\}$, another solution can be found by solving the second set of equations~\eqref{anisomgdin} involving the functions $\{f,\rho^{\theta},p^{\theta}_{r},p^{\theta}_{\perp}\}\, ,$ this is, three equations with four unknowns. In order to completely determine the system, extra conditions have to be implemented. Some of the cases are listed below:
\begin{itemize}
\item \textit{Interior solutions}. In this case, the mimic constraint for the radial pressure as an extra condition, namely $p=p^{\theta}_{r}$, have been used~\cite{Ovalle:2017wqi,Contreras:2019iwm,Torres:2019mee,Casadio:2019usg}.

\item \textit{Hairy Black Hole}. In this case, it is usual to impose suitable Equations of States (EoS) in the anisotropic sector~\cite{Ovalle:2018umz,Contreras:2018nfg,Rincon:2019jal}.

\item \textit{Inverse problem}. The constraint is
simply $\tilde{p}_{\perp}-\tilde{p}_{r}=p^{\theta}_{\perp}-p^{\theta}_{r}$, where $\tilde{p}_{\perp},\tilde{p}_{r}$ corresponds to the components of $T^{tot}_{\mu\nu}$. In contrast to the standard procedure, in this case it is assumed that a solution of Eq.~\eqref{einsorig} is given, and the goal is to explore both, the isotropic and decoupling sector. This problem has been worked out in $(3+1)$ and $(2+1)$-dimensions~\cite{Contreras:2018gzd}.
\end{itemize}

We note that once the system~\eqref{anisomgdin} is solved, the solution of Eq.~\eqref{einsorig} is given by $\{\nu,\lambda,\tilde{\rho},\tilde{p}_{r},\tilde{p}_{\perp}\}$, where $\lambda$ can be determined by using both, the decoupling equation~\eqref{decomposition} and the total energy-momentum tensor~\eqref{totalemt}.

Before concluding this Section, we would like to
point out that the decoupling source $\theta_{\mu\nu}$ could represent the coupling with scalar or vector fields \cite{Ovalle:2017fgl,Ovalle:2018ans}. Even more, this new matter content could encode the information of a new gravitational sector $X$
of extended theories of gravitation, whose Modified Einstein-Hilbert action $S_{MEH}$ can be expressed as \cite{Ovalle:2019qyi}
\begin{eqnarray}
S_{MEH}=S_{EH}+\int d^{4}x\sqrt{-g}\mathcal{L}_{X},
\end{eqnarray}
where $S_{EH}$ is the standard Einstein-Hilbert action\textbf{,} and $\mathcal{L}_{X}$ is the Lagrangian density of the 
$X-$gravitational sector, which can be written in terms of $\theta_{\mu\nu}$ as follows
\begin{eqnarray}
\theta_{\mu\nu}=\frac{2}{\sqrt{-g}}\frac{\delta \sqrt{-g}\mathcal{L}_{X}}{\delta g^{\mu\nu}}=2\frac{\delta\mathcal{L}_{X}}{\delta g^{\mu\nu}}-g_{\mu\nu}\mathcal{L}_{X}.
\end{eqnarray}

Such sector can be given by theories beyond general relativity, such as $f(R)$, Lovelock gravity \cite{Estrada:2019aeh}, Einstein-Aether gravity, among others \cite{Ovalle:2019qyi}.

\section{Gravitational decoupling for a Friedmann-Robertson-Walker spacetime}\label{FRW}

With the aim of determine the anisotropic term arising in a cosmological context, in this Section we implement the MGD approach to decouple an anisotropic metric from an isotropic sector given by the well known 
Friedmann-Robertson-Walker (FRW) metric, which in spherical coordinates is given by
\begin{eqnarray}
ds^{2}=-dt^{2}+\frac{a^{2}(t)}{1-k r^{2}}dr^2+a^{2}(t)r^{2}d\Omega^{2}\, .
\end{eqnarray}

Such line element can be written as
\begin{eqnarray}\label{pc}
ds^{2}=-e^{\nu}dt^{2}+\frac{dr^{2}}{\mu(r,t)}+R^{2}(r,t)d\Omega^{2},
\end{eqnarray}
where
\begin{subequations}
\begin{eqnarray}
e^{\nu}&=&1\, ,\\
\mu^{-1}(r,t)&=&\frac{a^{2}(t)}{1-k r^{2}}\, ,\label{muGR}\\
R(r,t)&=&a(t)r\, .
\end{eqnarray}
\end{subequations}

Note that with the parameterization~\eqref{pc}, the FRW metric looks formally like the line element of the isotropic sector in Eq.~\eqref{lei}, which is used as a seed for the MGD approach described in the previous Section. However, in contrast with Eq.~\eqref{lei}, the metric functions~\eqref{pc} depend also on the cosmic time $t$ through the scale factor $a(t)$. This feature leads to a system of differential equations arising from Eq.~\eqref{decomposition},\eqref{lel} and \eqref{einsorig} that can not be successfully decoupled if we insist in a geometric deformation with the form given by Eq.~\eqref{decomposition}, namely $\mu\to e^{-\lambda}=\mu+\alpha f$, and where the functions $\mu$ and $f$ depend only on the radial coordinate.
 
In order to overcome the difficulty mentioned above, we reformulate the MGD by proposing the following change: instead of considering the deformation given by Eq.~\eqref{decomposition}, we will consider
a more general transformation
\begin{eqnarray}
\mu\to e^{-\lambda}=\tilde{\mu}(t,r)\, ,
\end{eqnarray}
where $\tilde{\mu}$ contains the information from the isotropic sector through the function $\mu$ given by Eq.~\eqref{muGR}, and from the decoupling function $f$, which in general will be a function of both, the radial coordinate $r$ and the cosmic time $t$. After exploring several ways to include a geometric deformation, we found that in this case a suitable choice for $\tilde{\mu}$ is
\begin{eqnarray}
\tilde{\mu}(r,t)=\frac{a^{2}(t)}{1-kr^{2}+\alpha f(t,r)}\, ,
\end{eqnarray}
and then, Eq.~\eqref{lel} reads
\begin{eqnarray}\label{le}
ds^{2}=-dt^{2}+\frac{a^{2}(t)}{1-kr^{2}+\alpha f(t,r)}dr^{2}+a^{2}(t)r^{2}d\Omega^{2}\, ,
\end{eqnarray}
where it can be seen that the FRW line element is recovered in the limit $\alpha \rightarrow 0$. From now on, the implementation of the MGD
is straightforward. Considering Eq.~\eqref{le} as a solution of the EFE, we obtain
 \begin{subequations}\label{efefrw}
\begin{eqnarray}
G_{00} &=& 3\left[\left(\frac{\dot{a}}{a}\right)^2 + \frac{k}{a^2}\right] \nonumber \\
&-& \alpha \left[\frac{f+rf^{\prime}}{r^2a^2} + \left(\frac{\dot{a}}{a}\right)\frac{\dot{f}}{1 - kr^2 + \alpha f}\right] = \kappa^{2} \tilde{\rho}\, , \\
G_{01} &=& -\frac{\alpha \dot{f}(t)}{r\left[1 - kr^2 + \alpha f(t)\right]} = 0\, ,\label{G01} \\
G_{11} &=& -\frac{a^2}{(1 - kr^2 + \alpha f)}\left[\left( \frac{\dot{a}^2}{a^2} + 2\frac{\ddot{a}}{a} + \frac{k}{a^2} - \alpha \frac{f}{r^2a^2} \right)  \right] \nonumber \\
&=& \kappa^{2} \frac{\tilde{p}_ra^2}{1 + kr^2 + \alpha f(t)}\, , \\
G_{22} &=& -r^2\dot{a}^2 - 2r^2a\ddot{a} - r^2k + \frac{r\alpha f^{\prime}}{2} \nonumber \\
&+& \alpha r^2a\frac{2(1 - kr^2 + \alpha f)(a \ddot{f}+3\dot{a}\dot{f}) + 3\alpha a \dot{f}^2}{4\left[ 1 - 2kr^2 + k^2r^4 + \alpha f(2 - 2kr^2 + \alpha f) \right]} \nonumber \\
&=& \kappa^{2} \tilde{p}_{\perp}a^2r^2\, ,
\end{eqnarray}
\end{subequations}
where dots and primes denote derivatives with respect to cosmic time $t$ and radial coordinate $r$ respectively. According to Eq.~\eqref{totalemt}, we define
\begin{equation}
    \tilde{\rho}=\rho+\alpha\rho^{\theta}\, ,\quad \tilde{p}_{r}=p+\alpha p_{r}^{\theta}\, ,\quad \tilde{p}_{\perp}=p+\alpha p_{\perp}^{\theta}\, .
    \label{tildes}
\end{equation}

Notice that Eq.~\eqref{G01} imposes a constraint on $f$ given by $\dot{f}=0 \Rightarrow f(r,t) = f(r)$. Then, the EFE~\eqref{efefrw} reduce to
\begin{subequations}
\begin{eqnarray}
3\left[\frac{\dot{a}^2}{a^2} + \frac{k}{a^2}\right] - \alpha \left(\frac{f+rf^{\prime}}{r^2a^2}\right) &=& \kappa^{2} \tilde{\rho}\, , \label{eins1}\\
-\left[\left(\frac{\dot{a}}{a}\right)^2 + 2\frac{\ddot{a}(t)}{a(t)} + \frac{k}{a^2}\right] + \alpha \frac{f}{r^2a^2} &=& \kappa^{2} \tilde{p}_r\, , \label{eins2}\\
-\left[\left(\frac{\dot{a}}{a}\right)^2 + 2\frac{\ddot{a}}{a} + \frac{k}{a^2}\right] + \alpha \frac{ f^{\prime}}{2ra^2} &=& \kappa^{2} \tilde{p}_{\perp}\, . \label{eins3}
\end{eqnarray}
\end{subequations}

The above equations can be rewritten in terms of
two sets of differential equations: one describing an isotropic system sourced by the perfect fluid $T^{\mu}_{\nu}$, and the other
set corresponding to a new set of equations sourced by $\theta_{\mu\nu}$. Thus, for the perfect fluid we have
\begin{subequations}\label{isofrw}
\begin{eqnarray}
H^2 &=& \frac{\kappa^2}{3}\rho - \frac{k}{a^2}\, , \label{iso1}\\
\dot{H} &=& -\frac{\kappa^2}{2}\left( \rho + p \right) + \frac{k}{a^2}\, , \label{iso2}
\end{eqnarray}
\end{subequations}
where $H=\dot{a}/a$ is the Hubble parameter, and
\begin{subequations}\label{aniso}
\begin{eqnarray}
-\frac{r f'+f}{a^{2}r^{2}}&=&\kappa ^2  \rho^{\theta}\label{aniso1},\\
\frac{f}{a^2 r^2}&=&\kappa ^2 p^{\theta}_{r}\label{aniso2},\\
\frac{f'}{2 a^2 r}&=&\kappa ^2p^{\theta}_{\perp},\label{aniso3}
\end{eqnarray}
\end{subequations}
for the anisotropic system. It is
worth mentioning that Eq.~\eqref{aniso} induces a matter content for the anisotropic sector satisfying the following EoS
\begin{eqnarray}
p_{tot}^{\theta}=-\rho^{\theta}\, ,
\label{eos}
\end{eqnarray}
where $p_{tot}^{\theta} = p_{r}^{\theta}+2p_{\perp}^{\theta}$. On the other hand, the conservation equations~\eqref{conseq} lead to
\begin{subequations}
\begin{eqnarray}
0 &=& \dot{\rho} + 3H(\rho + p)\, ,\label{constmn} \\
0 &=& \dot{\rho}^{\theta} + H(3\rho^{\theta} + p_{tot}^{\theta})\, .
\label{conspimn}
\end{eqnarray}
\end{subequations}

Whereas Eq.~\eqref{constmn} will specify the ordinary matter content once given a EoS for a specific fluid (dust, radiation, etc), the combination of Eq.~\eqref{eos} and \eqref{conspimn} leads to an anisotropic energy density $\rho^{\theta}$ with form
\begin{equation}
    \rho^{\theta} = \frac{\rho_0^{\theta}}{a^2}\, ,
    \label{rhotheta}
\end{equation}
with $\rho_0^{\theta}$ the current value of the anisotropic energy density. The above expression allow us to find the decoupling function $f$ by integration of Eq.~\eqref{aniso1}, from where we obtain
\begin{equation}
    f(r) = - \frac{\kappa^2 \rho_0^{\theta}}{3}r^2\, ,
    \label{fr}
\end{equation}
where we have set the integration constant in such way that $f(0)=0$. The previous expression for $f(r)$ leads to the following radial and perpendicular components for the anisotropic pressure
\begin{equation}
    p^{\theta}_r = p^{\theta}_{\perp} = -\frac{1}{3}\rho^{\theta}\, ,
    \label{anipre}
\end{equation}

which clearly satisfies the EoS~\eqref{eos}. We can see that the anisotropic energy density~\eqref{rhotheta} has the cosmological evolution of a spatial curvature term. Therefore, the anisotropic sector $\theta_{\mu \nu}$ will contribute to the spatial geometry of the Universe. 

Notice that for a general, but constant EoS $\omega$, the energy density for different components of the Universe evolve as function of the scale factor $a$ as $\rho_i = \rho_{0,i}a^{-3(\omega+1)}$, where $i$ labels different matter contents each one with a characteristic EoS, for example baryons and CDM ($\omega=0$), photons and neutrinos ($\omega=1/3$), cosmological constant ($\omega=-1$), and spatial curvature ($\omega=-1/3$), which is precisely the relationship between the radial and perpendicular pressures with the anisotropic energy density, as we show above in Eq.~\eqref{anipre}. This result is a consequence of the constraint~\eqref{G01}, which forces to the decoupling function $f$ to be independent of the cosmic time $t$. This imposes the unique form of the EoS~\eqref{eos}, which combined with Eq.~\eqref{conspimn} leads to the evolution of the anisotropic energy density given by Eq.~\eqref{rhotheta}.

This is the case for a FRW metric, where the anisotropic sector $\theta_{\mu \nu}$ has the specific behavior of a spatial curvature term. Nonetheless, it could be different for other spacetimes. In the next Section, we will consider another line element, which will lead to different behaviors of the anisotropic sector.

\section{Gravitational decoupling for Kantowski-Sachs cosmology}\label{KS}
Now that we have studied the MGD in the standard FRW spacetime, and the conservation of the energy-momentum tensor leads to an unique and specific form of the decoupling function $f$, we implement the gravitational decoupling formalism for a Kanstowski-Sachs (KS) cosmology~\cite{Kantowski:1966te}. As we will see, for this metric it will be possible to induce different kind of fluids through the MGD. Let us start with the KS line element parameterized as 
\begin{equation}\label{kns}
ds^{2}=-dt^{2}+F(t)^{2}dr^{2}+S(t)^{2}d\Omega^{2}\, .
\end{equation}
Note that the KS metric can be written as Eq.~\eqref{lel}, whenever we identify
\begin{equation}
e^{-\lambda}\to F(t)^{2}\, ,\qquad r\to S(t)\, .
\end{equation}

Using Eq.~\eqref{kns}, the EFE \eqref{einsorig} take the form
\begin{subequations}
\begin{eqnarray}
\kappa^2\tilde\rho&=&\frac{2 S \dot{F} \dot{S}+F \dot{S}^2+F}{ F S^2}\, ,\\
\kappa^2\tilde{p}_{r}&=&-\frac{2 S \ddot{S}+\dot{S}^2+1}{S^2}\, ,\\
\kappa^2\tilde{p}_{\perp}&=&-\frac{S \ddot{F}+\dot{F} \dot{S}+F \ddot{S}}{F S}\, ,
\end{eqnarray}
\end{subequations}
where $\tilde{\rho}$, $\tilde{p}_{r}$, $\tilde{p}_{\perp}$ correspond to the components of the total energy-momentum tensor $T^{\mu\ (tot)}_{\nu}$ given by Eq.~\eqref{total}, and which can be splitted as Eq.~\eqref{totalemt}. Following the MGD approach, the standard KS matter content $T^{\mu}_{\nu}=diag(-\rho,p_{r},p_{\perp},p_{\perp})$ will be the source of a well-known solution of the KS metric, which is given by
\begin{eqnarray}\label{knsi}
ds^{2}=-dt^{2}+R(t)^{2}dr^{2}+S(t)^{2}d\Omega^{2}\, .
\end{eqnarray}

Following the strategy of the previous Section to decouple the EFE,
we assume that the $g^{rr}$ component of the line elements \eqref{kns} and \eqref{knsi} are related by
\begin{eqnarray}\label{decks}
F^{2}\to\frac{R^{2}}{1+\alpha f}\, ,
\end{eqnarray}
with $f$ the decoupling function. It is worth mentioning that Eq.~\eqref{decks} is the simplest deformation we can assume in order to be able to decouple the EFE. Note that, since the gravitational potentials of the KS line element do not depend on the radial coordinate $r$, the decoupling function $f$ can be set to be a function only of the cosmic time $t$. This is different from the FRW case, where the general assumption was $f(t,r)$. Thus, the EFE can be rewritten in two sets of equations: one set corresponding to the well-known KS solution
\begin{subequations}\label{ksiso}
\begin{eqnarray}
\kappa^2\rho &=& H_S^2 + 2H_SH_R + \frac{1}{S^2}\, ,\\
\kappa^2p_r &=&-\left(2\dot{H}_S + 3H_S + \frac{1}{S^2}\right)\, ,\\
\kappa^2p_{\perp}&=& -\left( \dot{H}_S + \dot{H}_R + H_S + H_R + H_SH_R \right)\, ,
\end{eqnarray}
\end{subequations}
where we have defined $H_S\equiv \dot{S}/S$ and $H_R\equiv \dot{R}/R\, ,$ and the other set containing the information of the decoupling sector
\begin{subequations}\label{anithetaks}
\begin{eqnarray}
\kappa^2 \rho^{\theta}&=&-\frac{\dot{f}H_S}{(1+\alpha  f)}\, ,\label{rhoksmgd}\\
\kappa^2 p_{r}^{\theta}&=&0\, , \label{pr0ks}\\
\kappa^2 p_{\perp}^{\theta}&=&
\frac{\ddot{f}}{2(1+\alpha  f)}
 -\frac{3 \alpha \dot{f}^{2}}{4(1+\alpha  f)^{2}}\, \nonumber\\
&&+\frac{\dot{f}}
{2(1+\alpha  f)}\left(2H_R+H_S\right)\, ,\label{pksmgd}
\end{eqnarray}
\end{subequations}
where we observe that there will be not contribution from the radial component of the anisotropic pressure. In fact, this behaviour coincides formally with the matter sector of the Florides interior solution~\cite{florides}, which represents an anisotropic Schwarzschild interior solution with vanishing radial pressure. This is quite interesting since the KS metric possess a very well-known property under the interchange $r\leftrightarrow t$, which maps from a Schwarzschild interior solution to an anisotropic (KS) cosmological solution, and viceversa~\cite{Kantowski:1966te} (some applications of this property in different contexts can be found at~\cite{Bronnikov:2003yi,LopezDominguez:2006wd,Bastos:2010zc,Djordjevic:2015uga}). Therefore, Eq.~\eqref{anithetaks} can be interpreted as the mapping of the anisotropic interior Schwarzschild solution~\cite{florides} to a cosmological KS background in presence of a MGD.

On the other hand, the conservation of the energy-momentum tensor leads to
\begin{subequations}
\begin{eqnarray}
0 &=& \dot{\rho} + H_R\left(3\rho + p_r\right) + 2H_S p_{\perp}\, ,\label{ksemtiso}\\
0 &=& \dot{\rho}^{\theta} + 3H_R\rho^{\theta} + 2H_S p_{\perp}^{\theta}- \frac{3\alpha \dot{f}}{2(1 + \alpha f)}\rho^{\theta}\, .\label{consks}
\end{eqnarray}
\end{subequations}

Contrary to the FRW case, where the conservation of $\theta_{\mu \nu}$ led to an unique expression for the anisotropic energy density (see Eq.~\eqref{rhotheta}), this time we need to provide an EoS in order to solve the system~\eqref{anithetaks}. Below we show the most general solutions for some particular cases of interest for several matter content:

\subsection*{Dust}

Let us impose the dust condition, namely
\begin{equation}
    p^{\theta}_{r}=p^{\theta}_{\perp}=0\, ,
\end{equation}
from where~Eq.\eqref{pksmgd} leads to
\begin{equation}
f(t)= \frac{4}{\alpha ^3 \left(-2 \sqrt{2} c_1\int \frac{dt}{R^2(t) S(t)} \, +c_2\right){}^2}-\frac{1}{\alpha }\, .\label{fdust}
\end{equation}

Now, replacing the above result into Eq.~\eqref{rhoksmgd}, the density for the decoupling sector will have the form
\begin{equation}
\rho^{\theta}=-\frac{\sqrt{2} c_1 H_S}{2 \pi  \alpha  R^2 S \left(c_2-2 \sqrt{2} c_1 \int \frac{dt}{R^2(t) S(t)} \, \right)}\, .\label{rhodust}
\end{equation}

\subsection*{Barotropic fluid}

Another possibility is to consider a barotropic EoS
\begin{equation}
    p_{\perp}^{\theta}=\omega \rho^{\theta}\, .
\end{equation}

Then, the geometric decoupling function $f$ and the anisotropic energy density $\rho^{\theta}$ are respectively given by
\begin{subequations}
\begin{eqnarray}
f(t)&=& \frac{4}{\alpha ^3 \left(c_2-2 \sqrt{2} c_1 \int \frac{S(t)^{-2 \omega-1}}{R^2(t)} \, dt\right){}^2}-\frac{1}{\alpha }\, , \label{fbaro}\\
\rho^{\theta}&=&\frac{c_1 S^{-2 (\omega+1)} \dot{S}}{\pi  \alpha  R^2 \left(4 c_1 \int \frac{S(t)^{-2 \omega-1}}{R^2(t)} \, dt-\sqrt{2} c_2\right)}\, .\label{4c1}
\end{eqnarray}
\end{subequations}

\subsection*{Politropic fluid}

A more interesting case arises when considering a polytropic fluid, which EoS have the form
\begin{equation}
    p^{\theta}_{\perp}=\omega \left(\rho^{\theta}\right)^{\beta}\, ,
\end{equation}
where $\beta=(n+1)/n$. For ultracompact objects, such a neutron stars, the polytropic index $n$ takes values from 0.5 to 1 for stiff EoS, or $n=1.5, 2$ for softer ones (see for instance~\cite{Baumgarte:1997eg,Hinderer:2007mb,Flanagan:2007ix}). In a cosmological context, several scenarios have been tested for polytropic fluid, from primordial to late time Universe~\cite{Adhav:2011zzi,BisnovatyiKogan:2011aa,Chavanis:2012pd,Chavanis:2012kla,Chavanis:2012uq,Freitas:2013nxa,Oztas:2018hoe,Setare:2018vhi}. When considering the case $n=1$ we have
\begin{equation}
f(t)=-\frac{1}{\alpha}\left[1-e^{\mathcal{H}(t)}\right]\, ,
\label{fpoli}
\end{equation}
where
\begin{equation}
\mathcal{H}=\int \frac{1}{R(t)^2 S(t) \left(c_1-\int^t \frac{\omega^2 S'(\tilde{t})^2+2 \pi  \alpha  S(\tilde{t})^2}{4 \pi  R(\tilde{t})^2 S(\tilde{t})^3} \, d\tilde{t}\right)} \, dt\, .
\end{equation}

In all the previous expressions, $c_1$ and $c_2$ are integration constants. It is straightforward to see that the general cases for the barotropic and polytropic solutions recover the most simple case of dust when $\omega=0$.

\subsection*{Cold Dark Matter}

The standard KS cosmology requires $p_r = p_{\perp}=0$ in order to have a CDM component, which evolution is given by $\rho_{CDM} = \rho_{CDM,0}/R^{3}$ after solving Eq.~\eqref{ksemtiso}. In our case for the decoupling sector, the condition $p_r^{\theta}=0$ is automatically satified by Eq.~\eqref{pr0ks}, and we have only to impose that $p_{\perp}^{\theta} = \dot{f}=0$, i.e., a constant decoupling function $f$ will contribute as cold dark matter, as can be seen from Eq.~\eqref{consks}. The dust condition applies for cold dark matter, in whose case we can ask for $f$ in Eq.~\eqref{fdust} to be constant by making the integration constant $c_1=0$. Nonetheless, this will lead to a null anisotropic energy density in Eq.~\eqref{rhodust}, or as can be seen directly from Eq.~\eqref{rhoksmgd}. The solution to obtain $\rho^{\theta}_{CDM}=\rho_{CDM,0}^{\theta}/R^3$ from the dust condition is in fact a more complicated one, in which an integro-differential equation relating the gravitational potentials $R$ and $S$ is obtained
\begin{equation}
    \frac{dS(t)}{dt} + \left( \beta_1 + \beta_2\int \frac{dt}{R^2(t)S(t)} \right)\frac{S^2(t)}{R(t)} = 0\, ,\label{idfcdm}
\end{equation}
where $\beta_1$ and $\beta_2$ are constants. Then, given solutions $R(t)$ and $S(t)$ from the standard KS sector satisfying the above equation, we can insert them into Eq.~\eqref{pksmgd} with the condition $p_{\perp}^{\theta}=0$, which leads to the following second order differential equation
\begin{equation}
\frac{\ddot{f}}{2(1+\alpha  f)}
 -\frac{\dot{f}}
{2(1+\alpha  f)}\left[\frac{3 \alpha \dot{f}}{2(1+\alpha  f)}-\left(2H_R+H_S\right)\right] = 0\, ,
\end{equation}
and thus, it is possible to determine the decoupling function $f$. Therefore, for a given solution $R(t)$ and $S(t)$ from the standard KS sector satisfying Eq.~\eqref{idfcdm}, the anisotropic energy density $\rho^{\theta}$ will behave as a cold dark matter component.

\subsection*{Cosmological Constant}

Different from the FRW case, the MGD for the KS cosmology allow us to propose several EoS, as we have shown above. An interesting question could be: is there a decoupling function $f$ capable of inducing a source with a behavior such as that of a cosmological constant? In other words, how should it be $f$ such that $p^{\theta}_{\perp} = -\rho^{\theta} = -\Lambda/\kappa^2$? The latter condition can be replaced in Eq.~\eqref{consks}, from where we obtain
\begin{equation}
    f(t) = -\frac{1}{\alpha}\left[ 1 - c_1 e^{\int \left[2H_R(t) - (4/3)H_S(t)\right]dt} \right]\, .
    \label{fcc}
\end{equation}

Therefore, the above expression leads to a Kantowski-Sachs Universe with cosmological constant by means of a purely geometric source.

\section{Discussions and Final Remarks}\label{remarks}
In this work we were able to find new general, analytical and exact cosmological solutions to the Einstein field equations by applying the Minimal Geometric Deformation approach. Particularly, we were focused on two cosmological metrics: the Friedmann-Robertson-Walker spacetime and a Kantowski-Sachs Universe, where an ansatz for the MGD in the spatial part of both metrics were proposed. This allowed us to use the Gravitational Decoupling formalism to find anisotropic extensions to the well-known solutions of these cosmological scenarios. Specifically, the set of Eqs.~\eqref{fr},~\eqref{fdust},~\eqref{fbaro},~\eqref{fpoli},~\eqref{fcc} constitutes different realizations of the decoupling function $f$, which induces a new gravitational sector. The major physical implications of this analysis are the following:\\

\textbf{1) Extended $\Lambda$CDM model}: the spatial curvature term in the FRW line element gets modified as
\begin{equation}
    [1-kr^2]^{-1} \rightarrow [1-kr^2+\alpha f(r)]^{-1}\, ,
\end{equation}
with $f(r)$ given by $f(r)=-(\kappa^2 \rho_{0}^{\theta}/3)r^2$ according to Eq.~\eqref{fr}. Therefore, an \textit{effective} spatial curvature term can be defined as
\begin{equation}
k\rightarrow k_{eff} \equiv k +  \alpha \frac{\kappa^2 \rho_{0}^{\theta}}{3}\, ,
\label{keff}
\end{equation}

\begin{center}
\begin{figure}[htp!]
    \centering
   \includegraphics[width=1.0\linewidth]{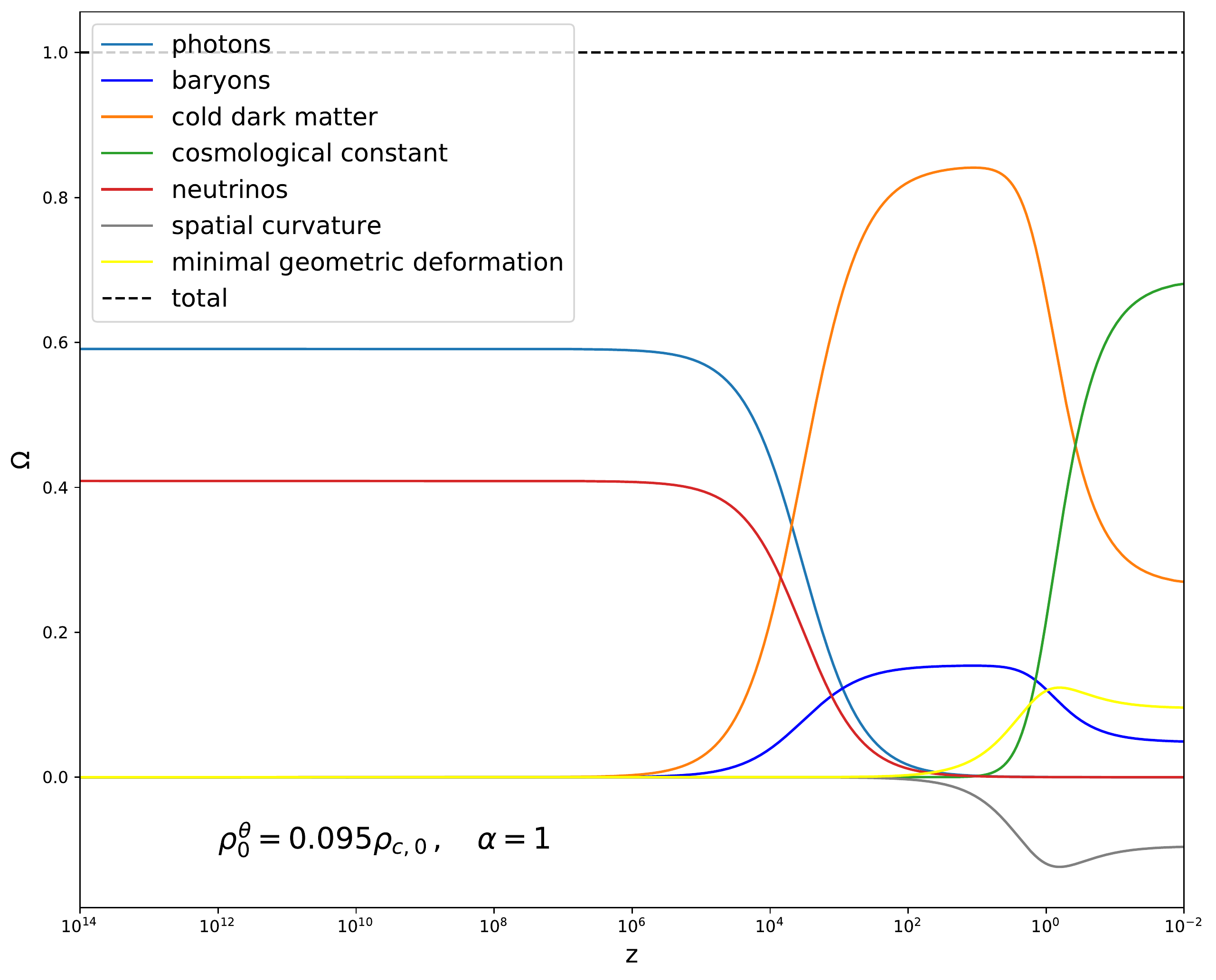}
        \label{fig:first_sub}   
    \caption{Cosmological evolution of the different components of the Universe. The effect of the anisotropic energy density (yellow line) is such that it balances the contribution of the spatial curvature parameter (gray line). These numerical solutions were obtained with an amended version of the \textsc{class} code~\cite{Lesgourgues:2011re}.}
    \label{plot1}
\end{figure}
\end{center}

Given the current values of the energy density parameters for both, total matter ($\Omega_{M,0}=0.315\pm 0.007$) and cosmological constant ($\Omega_{\Lambda,0}=0.685\pm 0.007$)~\cite{Aghanim:2018eyx}, Eq.~\eqref{keff} opens the possibility of a degeneration between the spatial curvature $k$ and the current value of the anisotropic energy density $\rho_0^{\theta}$, in such a way that 
\begin{equation}
    1 = \left( \Omega_{M,0} + \Omega_{\Lambda,0} + \Omega_{k_{eff},0} \right)\, ,\quad {\rm{with}}\quad  \Omega_{k_{eff},0}\simeq 0\, ,
\end{equation}
where we have defined
\begin{equation}\label{okeff}
    \Omega_{k_{eff}} \equiv \Omega_{k} + \alpha \Omega_{\theta }\, ,
\end{equation}
with $\Omega_{\theta }=\kappa^2\rho^{\theta}/3H^2$. Therefore, measurements of $\Omega_{k_{eff}}$ could be in fact indicating a non-flat spatial geometry in the FRW metric with $k=\pm 1$ balanced by the anisotropic term. While the standard $\Lambda$CDM cosmological model assumes a spatially flat geometry, the spatial curvature parameter $\Omega_k$ has been very well constrained from observations of the Cosmic Microwave Background (CMB) \cite{Ade:2015xua,Aghanim:2018eyx}, suggesting a nearly spatially flat Universe: $\Omega_{k}=0.0007\pm 0.0019$ (current constraint by Planck Collaboration 2018~\cite{Aghanim:2018eyx}). Assuming that it is the effective spatial curvature ${k_{eff}}$ which enters in the FRW metric, we can solve the background cosmological equations including the MGD contribution, i.e., the induced anisotropic energy density $\rho^{\theta}$ and pressure $p^{\theta}$ given by Eq.~\eqref{rhotheta} and Eq.~\eqref{anipre} respectively. Thus, we have
\begin{subequations}
\begin{eqnarray}
H^2 &=& \frac{\kappa^2}{3}\left( \sum_i \rho_i + \alpha \rho^{\theta} \right) - \frac{k}{a^2}\, ,\\
\dot{H} &=& -\frac{\kappa^2}{2}\left[ \sum_i \left(\rho_i + p_i\right) + \alpha \left(\rho^{\theta} + p^{\theta}\right)\right] + \frac{k}{a^2}\, ,\\
\dot{\rho}_i &+& 3H\left( \rho_i + p_i \right) = 0\, , \quad \dot{\rho}^{\theta} + 3H\left( \rho^{\theta} + p^{\theta} \right) = 0\, , \nonumber \\
\end{eqnarray}
\end{subequations}
where $i = \gamma\, , b\, , \nu\, , {\rm{cdm}}\, ,\Lambda\, ,$ stands for photons, baryons, neutrinos, cold dark matter, and cosmological constant respectively. It can be seen that $\alpha = 0$ recovers GR. In Figure~\ref{plot1} we show the cosmological evolution of each component of the Universe, from radiation era to the present.

At $z\sim 0$, we obtain the current values of the density parameters: $\Omega_{\gamma,0} = 5.419\times 10^{-5}\, , \Omega_{b,0} = 0.048\, , \Omega_{{\rm{CDM}},0} = 0.264\, , \Omega_{\Lambda,0} = 0.688\, , \Omega_{\nu,0} = 3.748\times 10^{-5}\, , \Omega_{k,0} = -0.095\, , \Omega_{\theta,0} = 0.095\, .$ As an example, we have chosen $\Omega_{k,0} = -0.095$, and the anisotropic energy density such that it totally balances the spatial curvature parameter (with $\alpha = 1$). Thus, we obtain $\Omega_{tot}=1$ during all the cosmological evolution (black dashed line).

We can go further, and calculate how much the MGD contributes to the spatial curvature. To do so, we perform a statistical analysis to compute the posteriors for the following set of parameters $\left\lbrace \Omega_k\, , \Omega_{\theta}\, , \alpha \right\rbrace$. Since we are focused in the background equations, we used the Planck Compressed 2018 data~\cite{Chen:2018dbv} instead of the full Planck likelihood 2018.

\begin{center}
\begin{figure}[htp!]
    \centering
   \includegraphics[width=1.0\linewidth]{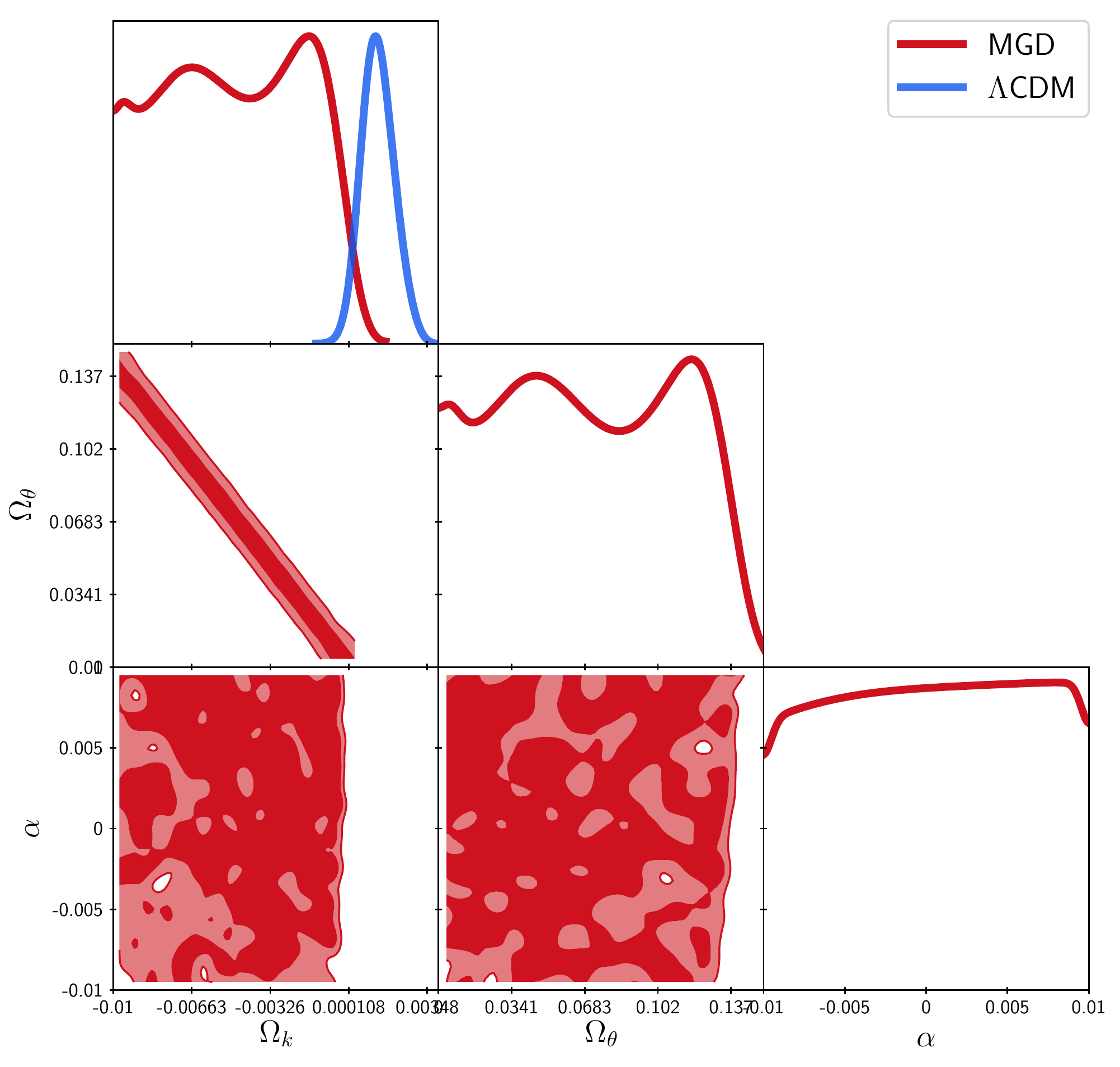}
        \label{fig:first_sub}   
    \caption{Posteriors for the cosmological parameters $\left\lbrace \Omega_k\, , \Omega_{\theta}\, , \alpha \right\rbrace $ using CMB data. See text for more details. The posteriors were computed with the cosmological parameter estimator \textsc{monte python}~\cite{Audren:2012wb}.}
    \label{plot2}
\end{figure}
\end{center}

After exploring several priors ranges, we finally consider flat priors within the following values: $\Omega_k = [-0.01,0.01]\, ,\Omega_{\theta} = [0,1]\, ,\alpha = [-0.01,0.01]$. When analyzing the chains, all the parameters satisfied the Gelman-Rubin criterium~\cite{Gelman:1992zz}, in particular $R-1<0.001$. The posteriors are shown in Figure~\ref{plot2}, where it can be seen how well constrained $\Omega_k$ is when inferred by CMB observations within the standard $\Lambda$CDM model plus spatial curvature: $\Omega_k = 0.00137^{+0.000703}_{-0.000691}$ (blue posterior).

However, we observe that the MGD contribution (red posteriors) allows the spatial curvature to have negative values. In fact, the parameters $\left\lbrace \Omega_k\, ,\Omega_{\theta} \right\rbrace$ are shown to be anti-correlated within the range $0\leq \Omega_{\theta} \leq 0.14$ and $-0.01\leq \Omega_{k} \leq 1.1\times 10^{-4}$, whereas the constant $\alpha$ measuring the strength of the geometric deformation is not constrained, and it presents a flat posterior. We want to emphasize that the behavior of these posteriors persist for broader priors. Even though the MGD energy density parameter can reach $\Omega_{\theta}\sim 10\%$ of the total matter-energy budget, it has to be recalled that it is weighted by the constant $\alpha$. Moreover, we analyze $\Omega_{k_{eff}}$ as a derived parameter (see Eq.~\eqref{okeff}), and the most likely values it can takes are given by $-6\times 10^{-4} > \Omega_{k_{eff}} > -10^{-2}$, implying a slight preference for a closed Universe, as was recently reported by~\cite{DiValentino:2019qzk}. However, other set of observations must be considered in our analysis to study how much the posteriors change, and whether $\Omega_{k_{eff}}$ turns out to be consistent with a flat Universe.

It would be interesting to explore the general geometric deformation given by Eq.~\eqref{gdt}. This could lead to new matter terms that may play a relevant cosmological role in the evolution of the FRW Universe. However, it will imply to find the appropriate way to introduce the $g$ function such that the decoupling of the EFE is possible. This has been computed, for instance, for astrophysical systems on Braneworld models~\cite{Ovalle:2016pwp}, and for the Einstein-Maxwell system as well~\cite{Ovalle:2019qyi}.

\textbf{2) CDM and $\Lambda$ from geometry}: Contrary to the FRW case, where the EoS sets the behavior of the anisotropic component in an unique form, we found that a Kantowski-Sachs Universe allows to have a variety of matter components induced from the decoupling sector. By specifying a particular solution for $S(t)\, ,R(t)$ in a standard KS model, its extended version with a new geometric component arising from the decoupling function $f$ can be computed. We showed that it is possible to map $f$ to matter-like terms, such as dust, barotropic and polytropic fluids, cold dark matter, and to a cosmological constant as well. Even when the most accepted cosmological model is $\Lambda$CDM, the Universe we observe is in a non-linear phase of structure formation and of accelerating expansion, features attributed to new forms of matter (Cold Dark Matter) and energy (Cosmological Constant). Then, we want to highlight the benefits of having the freedom of realising the description of a dark matter component or a cosmological constant in the Kantowski-Sachs universe, since these components are instead required in FRW universe to explain observations. Thus, in anisotropic Universes like those provided by Kantowski-Sachs models, in combination with the MGD, the cold dark matter fluid and the cosmological constant arise as geometric effects of the spacetime.

Therefore, the Gravitational Decoupling formalism through the Minimal Geometric Deformation approach promises to be useful to explore geometric aspects of the large scale Universe, and to incorporate matter and energy components that could drive the process of structures formation and late time acceleration as well. This will be interesting to analyze with some deep in future studies.

\section*{Acknowledgement}
FXLC is supported by a PRODEP postdoctoral fellowship at Instituto de F\'isica y Matem\'aticas in the Universidad Michoacana de San Nicol\'as de Hidalgo (IFM-UMSNH).

\bibliography{bib}

\end{document}